\documentclass [prc,aps,showpacs]{revtex4}
\usepackage[usenames]{color}
\usepackage{ulem} 

\usepackage{graphicx}

\begin{document}

\title{Caloric curve for nuclear liquid-gas phase transition in relativistic mean-field hadronic model}

\author{A.S.~Parvan}

\affiliation{Bogoliubov Laboratory of Theoretical Physics, Joint Institute for Nuclear Research, 141980 Dubna, Russian Federation}

\affiliation{Institute of Applied Physics, Moldova Academy of Sciences, MD-2028 Chisinau, Republic of Moldova}

\begin{abstract}
The main thermodynamical properties of the first order phase transition of the relativistic mean-field (RMF) hadronic model were explored in the isobaric, the canonical and the grand canonical ensembles on the basis of the method of the thermodynamical potentials and their first derivatives. It was proved that the first order phase transition of the RMF model is the liquid-gas type one associated with the Gibbs free energy $G$. The thermodynamical potential $G$ is the piecewise smooth function and its first order partial derivatives with respect to variables of state are the piecewise continuous functions. We have found that the energy in the caloric curve is discontinuous in the isobaric and the grand canonical ensembles at fixed values of the pressure and the chemical potential, respectively, and it is continuous, i.e. it has no plateau, in the canonical and microcanonical ensembles at fixed values of baryon density, while the baryon density in the isotherms is discontinuous in the isobaric and the canonical ensembles at fixed values of the temperature. The general criterion for the nuclear liquid-gas phase transition in the canonical ensemble was identified.
\end{abstract}

\pacs {21.65.-f; 05.70.Fh; 25.70.Mn}

\maketitle 

\section{Introduction}
The study of the phase transitions in the nuclear matter under extreme conditions is important for the theoretical analysis of heavy ion collision experiments at the intermediate and high energies~\cite{Yagi,Bondorf} and for understanding of many critical issues in astrophysics~\cite{Steiner05}. It is generally accepted that the nuclear multifragmentation observed in intermediate-energy nuclear reactions is indicating the first order phase transition of the nuclear liquid-gas type~\cite{Mishustin06}. This claim is effectively based on the concept of the plateau in the nuclear multifragmentation caloric curve (the dependence of the temperature $T$ on the excitation energy $E^*$ of the system) which was theoretically predicted in \cite{Bondorf85} and later was experimentally found in the $600$ MeV/nucleon $Au+Au$ collision~\cite{Pochodzalla}. However, in certain experiments, as for example in  $1$ GeV/nucleon $Au+C$ collision \cite{Hauger} and $95$ MeV/nucleon $Ar+Ni$ collisions \cite{Ma97}, the plateau in the experimental caloric curve is absent. Moreover, although there is a class of statistical multifragmentation models (SMM) which predicts a plateau (the discontinuity of energy) in the caloric curves in the canonical and microcanonical ensembles \cite{Bondorf,Bondorf85,Bondorf98,Gross90,Das2003,Raduta02,Scharenberg01,Parvan00}, there is yet another class of SMM which predicts a plateau in the caloric curves only in the isobaric ensemble and a very smooth transition in the canonical and microcanonical ensembles \cite{Elliott00,Samaddar04,Aguiar06,De07,DasGupta01,Parvan99}. This particular property of the caloric curves for the first order phase transition in the canonical and microcanonical ensembles can be also seen in the models of other physical systems \cite{Campa09,Huller94,Ota01,Gross97,Lee96,Chomaz99,Chomaz02,Barre02,Mulken01}.

The first order phase transition in almost all classification schemes has the Ehrenfest definition, which is one associated with a finite discontinuity in one or more of the first derivatives of the appropriate thermodynamic potential with respect to its variables of state~\cite{Ehrenfest,Papon}. The concrete type of the first order phase transition is related to an appropriate potential~\cite{Yeomans}. For example, for the liquid-gas phase transition the Gibbs free energy $G$ is relevant and there are discontinuities in its first derivatives, the entropy and the volume, across the temperature-pressure coexistence curve~\cite{Papon}. Herewith, the energy in the caloric curve for the liquid-gas phase transition is discontinuous in the isobaric ensemble and it is continuous in the canonical and the microcanonical ensembles. Otherwise, for a magnetic system the free energy $F$ is the appropriate potential with the discontinuity in its first derivative, the magnetization, across the temperature-magnetic field coexistence curve~\cite{Yeomans}. The energy in the caloric curve for the first order phase transition associated with the free energy $F$ is discontinuous in the canonical and the microcanonical ensembles, as the first derivative of $F$, the entropy, is discontinuous~\cite{Stanley}. Some other alternative methods which deal with the occurrence of phase transitions can be found in~\cite{Landau,Yang,Fisher,Lee,Chomaz1,Chomaz2,Gross,Gorenstein,Bugaev}.

The nuclear liquid-gas phase transition, i.e. the first order phase transition associated with the Gibbs free energy $G$, usually occurs in the models of the interacting hadron gas such as the relativistic mean-field hadronic model~\cite{Walecka74,Serot86,Silva,Avancini04,Avancini06,Ducoin08,Ayik11}, the mean-field approximation with a Skyrme effective interaction~\cite{Ducoin06}, etc. In spite of the fact that these models have been intensively investigated during a long period of time, the liquid-gas phase transition in such models has not been studied completely. Therefore, the main purpose of this paper is to provide a holistic and concerted approach to tackle the problem of the nuclear liquid-gas phase transition on the basis of the method of the thermodynamical potentials and their first derivatives in different statistical ensembles and to establish the general criteria for the liquid-gas phase transition which allow to distinguish it from the first order phase transition associated with the free energy $F$. To determine these criteria we use the well-known RMF hadronic model~\cite{Walecka74,Serot86}, which enables to find the consistent behavior of the caloric curve and the equation of state (the discontinuity of the density in the isotherms~\cite{Goodman84,Huang}) for the nuclear liquid-gas phase transition in different statistical ensembles.

The structure of the paper is as follows. In Section~\ref{RMF}, we briefly describe basic ingredients of the relativistic mean-field model and the methodology to evaluate the model parameters. The thermodynamic results for the liquid-gas phase transition and the caloric curve are discussed in Sections~\ref{Rez1}. The main conclusions are summarized in the final section. 

\section{Relativistic mean-field hadronic model}\label{RMF}
Let us consider the relativistic hadron many-body problem in the framework of the local quantum field theory. In the Lagrange formalism this theory is based on the action functional of the form $\mathcal{S}=\int d^{4}x\mathcal{L}$, where $d^{4}x$ is the integration measure in four-dimensional Minkowski space and $\mathcal{L}$ is the Lagrange density. We now proceed to use the simplest realization of the RMF approach~\cite{Walecka74}. The effective Lagrangian of the system of the low-lying baryons interacting through the exchange of the scalar meson $\sigma$ and the intermediate abelian $U(1)$ gauge boson, i.e. vector meson $\omega$, is given by~\cite{Serot86}
\begin{equation}\label{1}
\mathcal{L} = \frac{\imath}{2} \left[\overline{\psi} \gamma^{\mu}
D_{\mu}\psi- (D_{\mu}^{*}\overline{\psi})
\gamma^{\mu}\psi \right] - m\overline{\psi}\psi  -  \frac{1}{4} F_{\mu\nu}F^{\mu\nu} +\frac{1}{2} m_{v}^{2}A_{\mu}A^{\mu} +
\frac{1}{2}(\partial_{\mu}\phi\partial^{\mu}\phi-m_{s}^{2}\phi^{2})+g_{s}\phi\overline{\psi}\psi,
\end{equation}
where
$
\psi=\left(\begin{array}{c}
  \psi^{1} \\
  \psi^{2} \\
\end{array}\right)
$
is the isodoublet of the group SU(2), $\psi^{1}$ and $\psi^{2}$ are the spinor fields for protons and neutrons with common mass $m$, $\gamma^{\mu}$ is the Dirac gamma matrix, $A^{\mu}$ is the real isoscalar vector Proca field for the $\omega$ meson with mass $m_{v}$, $\phi$ is the real isoscalar scalar field for $\sigma$ meson with mass $m_{s}$ and $g_{v},g_{s}$ are the coupling constants, respectively. The antisymmetric field strength tensor for gauge field and the covariant derivative of the group U(1) are defined by
\begin{equation}\label{2}
F_{\mu\nu} = \partial_{\mu} A_{\nu}-\partial_{\nu} A_{\mu}, \qquad D_{\mu} = \partial_{\mu}+\imath g_{v}A_{\mu}.
\end{equation}
Note that throughout the paper we use the natural units system, $\hbar=c=k_B=1$.

This Lagrangian resembles the quantum electrodynamics (QED) Lagrangian with a massive abelian "photon" and an additional scalar interaction. This theory is  renormalizable~\cite{Serot86,Okun}. The Lagrangian (\ref{1}) is invariant with respect to the global gauge transformation of the abelian group $U(1)$ and the non-abelian group $SU(2)$. But, it is not invariant with respect to the local gauge transformations of these two groups. The mass term $m_{v}^{2}A_{\mu}A^{\mu}$ in the Lagrangian (\ref{1}) destroys the local abelian gauge U(1) invariance. In order to be invariant with respect to the local gauge transformations of the non-abelian group SU(2), the Lagrangian (\ref{1}) must contain an isotopic triplet of massless vector fields interacting with the spinor fields $\psi^{1}$ and $\psi^{2}$, because the theory with non-abelian massive gauge fields is not renormalizable. If that is the case, to preserve the renormalizability of the Lagrangian, the mass terms for the non-abelian gauge quanta should be introduced into the theory via the so-called Higgs mechanism~\cite{Okun}. However, in isospin symmetric matter the long-range part of the nuclear force which comes from one pion exchange and the force mediated by the exchange of the rho meson average to zero.

In the mean-field approximation~\cite{Serot86}, the following conditions are imposed on the fields $A_{\mu}$ and $\phi$,
\begin{equation}\label{4}
\phi=\phi_0, \qquad  A_{\mu}= \delta_{\mu}^{0} A_{0},
\end{equation}
where the quantities $\phi_0$ and $A_0$ are constants, independent of $x_\mu$ (i.e. $\partial_{\mu}A_0=0$ and $\partial_{\mu}\phi_{0}=0$). In the second quantization, the energy operator, $\hat{H}=\int d^{3}x T^{00}$, the net baryon charge operator, $\hat{B}=\int d^{3}x \overline{\psi}\gamma^{0}\psi$, and the Euler-Lagrange equations for the classical fields $A_{0}$ and $\phi_{0}$ can be written as~\cite{Serot86}
\begin{eqnarray}\label{6}
\hat{H} &=& \sum\limits_{\vec{p}\sigma} \ [\varepsilon_{(+)}a_{\vec{p}\sigma}^{\dag}a_{\vec{p}\sigma}+\varepsilon_{(-)}
b_{\vec{p}\sigma}^{\dag}b_{\vec{p}\sigma}]- \frac{1}{2}(m_{v}^{2}A_{0}^{2}-m_{s}^{2}\phi_{0}^{2})V, \\ \label{7}
\hat{B} &=& \sum\limits_{\vec{p}\sigma} \ [a_{\vec{p}\sigma}^{\dag}a_{\vec{p}\sigma}- b_{\vec{p}\sigma}^{\dag}b_{\vec{p}\sigma}], \\ \label{8}
A_{0}&-&\frac{g_{v}}{m_{v}^{2}}\frac{1}{V}\sum\limits_{\vec{p}\sigma} [a_{\vec{p}\sigma}^{\dag}a_{\vec{p}\sigma}-
b_{\vec{p}\sigma}^{\dag}b_{\vec{p}\sigma}]=0, \\ \label{9}
\phi_{0} &-& \frac{g_{s}}{m_{s}^{2}}\frac{1}{V}\sum\limits_{\vec{p}\sigma}\frac{m^{*}}{E^{*}} \ [a_{\vec{p}\sigma}^{\dag}a_{\vec{p}\sigma}+
b_{\vec{p}\sigma}^{\dag}b_{\vec{p}\sigma}] = 0,
\end{eqnarray}
where $\varepsilon_{(\pm)}=E^{*}\pm g_{v}A_{0}$ is the zero component of the four momentum, $p_{(\pm)}^{\mu}=(\varepsilon_{(\pm)},\vec{p})$, of the plane-wave solutions of the Dirac equation, $E^{*}=\sqrt{\vec{p}^{2}+m^{*2}}$ is the zero component of the four momentum $p^{\mu}=(E^{*},\vec{p})$ of the Dirac bispinor of amplitudes $u_{\pm p,\pm \sigma}$, $m^{*}=m-g_{s}\phi_{0}$, is the nucleon effective mass, $p^{2}=m^{*2}$ and $V$ is the volume of the system. We use the composite index notation, $\sigma=(s,I)$, where $s$ and $I$ are the nucleon spin and isospin components, respectively. The creation and annihilation operators for fermions, $a_{\vec{p}\sigma}^{\dag}$ and $a_{\vec{p}\sigma}$, and anti-fermions, $b_{\vec{p}\sigma}$ and $b_{\vec{p}\sigma}^{\dag}$, satisfy the equal-time anticommutation relations. Note that in Eqs.~(\ref{6}) -- (\ref{9}) the vacuum terms were neglected.

Let us consider a system of volume $V$, in contact with a heat and particle reservoir of temperature $T$ and chemical potential $\mu$. Then, the statistical operator and partition function are
\begin{equation}\label{10}
\hat{\varrho}=\frac{1}{Z}\ e^{-\frac{\hat{H}-\mu\hat{B}}{T}}
\qquad {\rm and}\qquad
Z = Tr\left(e^{-\frac{\hat{H}-\mu\hat{B}}{T}}\right),
\end{equation}
respectively. Plugging Eqs. (\ref{6})-(\ref{9}) into (\ref{10}), we obtain~\cite{Serot86}
\begin{eqnarray}\label{20}
\Omega = -T\ln Z &=& - T \sum\limits_{\vec{p}\sigma} \left[\ln (1+e^{-\frac{\varepsilon_{(+)}-\mu}{T}})+\ln
 (1+e^{-\frac{\varepsilon_{(-)}+\mu}{T}}) \right]  -\lambda_{(-)} , \\ \label{12}
  \rho &-& \frac{1}{V} \sum\limits_{\vec{p}\sigma} [\langle n_{\vec{p}\sigma}\rangle - \langle \bar{n}_{\vec{p}\sigma}\rangle]=0,  \\ \label{13}
  \rho_{s} &-& \frac{1}{V} \sum\limits_{\vec{p}\sigma} \frac{m^{*}}{E^{*}}[\langle n_{\vec{p}\sigma}\rangle +
  \langle \bar{n}_{\vec{p}\sigma}\rangle]=0,
\end{eqnarray}
with the mean occupation numbers given by the Fermi-Dirac distribution functions
\begin{equation}\label{14}
    \langle n_{\vec{p}\sigma}\rangle = \frac{1}{e^{\frac{\varepsilon_{(+)}-\mu}{T}}+1}, \qquad
\langle \bar{n}_{\vec{p}\sigma}\rangle = \frac{1}{e^{\frac{\varepsilon_{(-)}+\mu}{T}}+1}
\end{equation}
and
\begin{equation}\label{15}
  \lambda_{(\mp)} = \frac{1}{2} (a_{v}\rho^{2}\mp a_{s}\rho_{s}^{2})V, \qquad \varepsilon_{(\pm)} = E^{*}\pm a_{v}\rho, \qquad m^{*} = m-a_{s}\rho_{s},
\end{equation}
where $\rho \equiv A_{0}g_{v}/a_{v}$, $\rho_{s} \equiv \phi_{0}g_{s}/a_{s}$, $a_{v}\equiv g_{v}^{2}/m_{v}^{2}\equiv C_{v}^{2}/m^2$, $a_{s} \equiv g_{s}^{2}/m_{s}^{2}\equiv C_{s}^{2}/m^2$, with $C_v$ and $C_s$ being dimensionless parameters of the model~\cite{Serot86}.

Equations (\ref{20})--(\ref{15}) constitute a closed system of equations which allow us to determine all ensemble averages. Using Eqs.~(\ref{20})-(\ref{13}), we observe that the thermodynamic potential is independent of $\rho$ and $\rho_{s}$:
\begin{eqnarray}\label{20a}
\left(\frac{\partial \Omega}{\partial\rho}\right)_{TV\mu\rho_{s}} &=&  - a_{v} V ( \rho  - \frac{1}{V} \sum\limits_{\vec{p}\sigma}
  [\langle n_{\vec{p}\sigma}\rangle - \langle \bar{n}_{\vec{p}\sigma}\rangle]) = 0,  \\ \label{20aa}
\left(\frac{\partial \Omega}{\partial\rho_{s}}\right)_{TV\mu\rho} &=&  a_{s} V ( \rho_{s}  - \frac{1}{V} \sum\limits_{\vec{p}\sigma}
  \frac{m^{*}}{E^{*}}[\langle n_{\vec{p}\sigma}\rangle + \langle \bar{n}_{\vec{p}\sigma}\rangle]) = 0.
\end{eqnarray}
In particular, the fundamental equations of thermodynamics are
\begin{eqnarray}\label{21a}
  d\Omega &=& -SdT-pdV-\langle B\rangle d\mu, \\ \label{21b}
TdS &=& d \langle H\rangle +pdV-\mu d\langle B\rangle
\end{eqnarray}
and the Euler theorem is
\begin{equation}\label{21c}
     TS = \langle H\rangle +pV-\mu\langle B\rangle.
\end{equation}
The mean energy $\langle\hat H\rangle$, the mean net baryon charge $\langle\hat B\rangle$, the pressure $p$, and the entropy $S$ can be written as
\begin{eqnarray}\label{21}
  \langle H\rangle  &=& -T^{2}\frac{\partial}{\partial T}\left(\frac{\Omega}{T}\right)_{V\mu}+\mu \langle B\rangle = \sum\limits_{\vec{p}\sigma}
  E^{*}[\langle n_{\vec{p}\sigma}\rangle + \langle \bar{n}_{\vec{p}\sigma}\rangle]+ \lambda_{(+)}, \\
\label{22}
\langle B\rangle &=& -\left(\frac{\partial\Omega}{\partial \mu}\right)_{TV} = \sum\limits_{\vec{p}\sigma}
  [\langle n_{\vec{p}\sigma}\rangle - \langle \bar{n}_{\vec{p}\sigma}\rangle], \\
\label{23}
 p &=& -\left(\frac{\partial \Omega}{\partial V}\right)_{T\mu} = \frac{1}{3V} \sum\limits_{\vec{p}\sigma}
 \frac{\vec{p}^{2}}{E^{*}}[\langle n_{\vec{p}\sigma}\rangle + \langle \bar{n}_{\vec{p}\sigma}\rangle]+\frac{\lambda_{(-)}}{V}, \\
\label{24}
S &=& -\left(\frac{\partial \Omega}{\partial T}\right)_{V\mu}= \frac{1}{T} (-\Omega+\langle H\rangle - \mu \langle B\rangle).
\end{eqnarray}
where $\rho=\langle B\rangle/V$ is the density of the net baryon charge.

Let us introduce the notation  $\varepsilon=\langle H\rangle/V$. The constants $a_{s}$ and $a_{v}$ are derived from the lowest-energy state of the system at $T=0$ and the conditions $\varepsilon_{b}(\rho_{0}) = \varepsilon_{b0}$ and $\rho=\rho_{0}$, where $\varepsilon_{b} = (\varepsilon/\rho)-m$ is the binding energy per nucleon and $\rho_{0}$ is the normal nuclear density. For the isospin symmetric nuclear matter $\varepsilon_{b0} = -16$ MeV and $\rho_{0} = 0.16$ fm$^{-3}$, respectively \cite{Serot86}. We obtain~\cite{Anghel}
\begin{eqnarray}
\label{33}
  a_{v} &=& \frac{1}{\rho_{0}} \left(\varepsilon_{b0}+m-\frac{p_{F0}}{x_{0}}\sqrt{1+x_{0}^{2}}\right), \\
\label{34}
  a_{s}^{-1} &=& \frac{3}{2}\rho_{0} \frac{I_{2}(x_{0})}{x_{0}^{3}}\left(m-\frac{p_{F0}}{x_{0}}\right)^{-1}, \\
\label{35}
0 &=& \frac{3}{4} p_{F0} \frac{I_{1}(x_{0})}{x_{0}^{4}} + \frac{3}{2} \frac{I_{2}(x_{0})}{x_{0}^{3}}\left(m-\frac{p_{F0}}{x_{0}}\right) - \varepsilon_{b0}- m -\frac{p_{F0}}{x_{0}}\sqrt{1+x_{0}^{2}}
\end{eqnarray}
and
\begin{equation}\label{30}
 b_{\nu} I_{\nu}(x) = x\sqrt{1+x^{2}}(1+c_{\nu} x^{2}) - \ln(x+\sqrt{1+x^{2}}),
\end{equation}
where $\nu=1,2,3$, $x_{0} =p_{F0}/m_{0}^{*}$, $m_{0}^{*}=m^{*}(\rho_{0})$, $p_{F0}=(6\pi^{2} \rho_{0}/\gamma)^{1/3}$ is the Fermi momentum, $\gamma$ is the spin-isospin degeneracy factor for nucleons, $b_{1}=b_{2}=1$, $b_{3}=-1$ and $c_{1}=2,c_{2}=0,c_{3}=-2/3$. The number $x_{0}$ (or $m_{0}^{*}$) is the solution of Eq.~(\ref{35}). Substituting $x_{0}$ (or $m_{0}^{*}$) into Eqs.~(\ref{33}) and (\ref{34}), we obtain the numerical values for the parameters of the model at $T=0$: $C_{v}^{2}\simeq 250$ and $C_{s}^{2}\simeq 330$.

\begin{figure}[t]
\includegraphics[width=14cm]{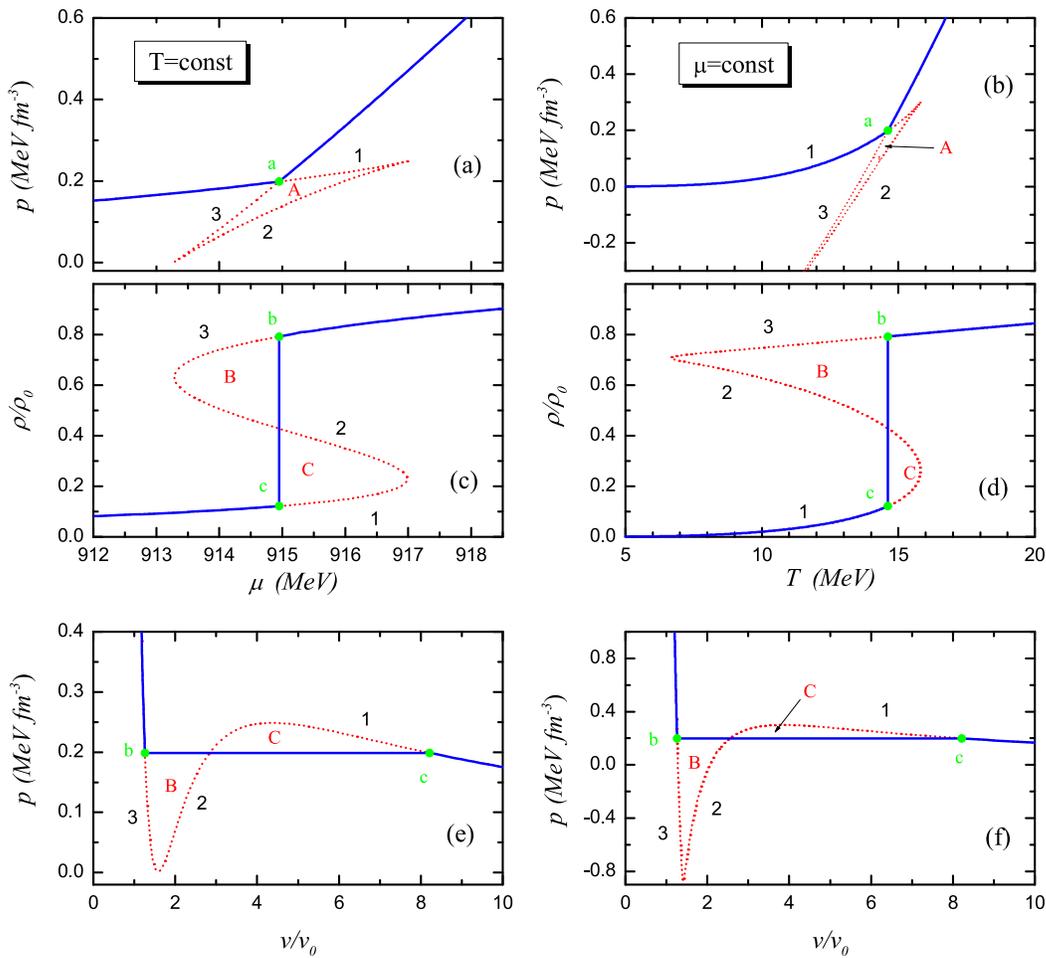} \vspace{-0.4cm}
\caption{(Color online) The Maxwell construction for the specific thermodynamical potential $p(T,\mu)$ (the pressure), its first order partial derivative, $\rho=\partial p/\partial \mu$ (upper part) and for the pressure $p$ as a function of the specific volume $v$ (lower part) of the RMF model in the grand canonical ensemble at fixed temperature $T$ (left panels) and at fixed chemical potential $\mu$ (right panels). The curves depict the exact anomalous results (dotted lines) and the results with the Maxwell construction (solid lines). Symbols denote the phase transition point at $T=14.62$ MeV and $\mu=914.95$ MeV. The normal specific volume $v_{0}=1/\rho_{0}=6.25$ fm$^{3}$, where $\rho_{0}$ is the normal nuclear density, $\rho_{0}=0.16$ fm$^{-3}$.} \label{a1}
\end{figure}

\section{First order phase transition for nuclear matter}\label{Rez1}
\subsection{Grand canonical ensemble $(T,V,\mu)$}

Let us describe the phase transition of the RMF model in the grand canonical ensemble. This ensemble is characterized by the potential, $\Omega(T,V,\mu)$ (\ref{20}) and the physical observables correspond to its first and second order partial derivatives, Eqs. (\ref{22}), (\ref{23}) and (\ref{24}). The geometry of the hypersurface $\Omega$ determines the physical properties of the system. The thermodynamical potential (\ref{20}) is a homogeneous function of first degree of the extensive variable of state $V$, so in the follwing we shall work with the specific thermodynamic potential $\omega$ (the pressure $p$),
\begin{equation}\label{36}
    \omega(T,\mu)\equiv \Omega(T,V,\mu)/V = -p(T,\mu).
\end{equation}
From (\ref{36}) we obtain the entropy and baryon densities,
\begin{equation}\label{37a}
   s(T,\mu)\equiv\frac{S}{V}=\left(\frac{\partial p}{\partial T} \right)_{\mu}, \qquad  \rho(T,\mu)\equiv\frac{\langle B\rangle}{V}=\left(\frac{\partial p}{\partial \mu} \right)_{T},
\end{equation}
which satisfy Eqs. (\ref{21a}), (\ref{21b}), and (\ref{21c}), as
\begin{equation}\label{37}
  dp = sdT+\rho d\mu, \qquad  Tds = d\varepsilon -\mu d\rho, \qquad  Ts=\varepsilon+p-\mu\rho.
\end{equation}

\begin{figure}[htbp]
\includegraphics[width=14.5cm]{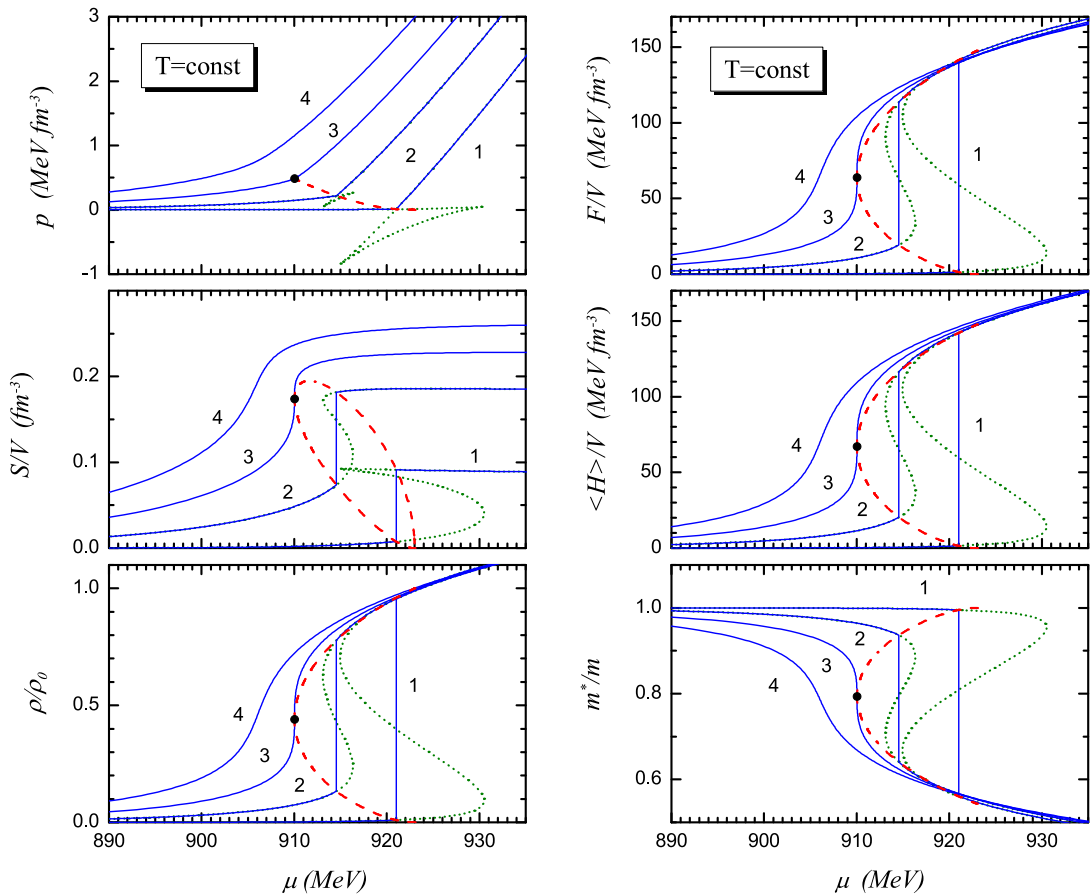} \vspace{-0.3cm}
\caption{(Color online) Grand canonical ensemble $(T,\mu)$. The specific thermodynamical potential $p$ (the pressure), the entropy density $S/V$, the baryon density $\rho$, the density of free energy $F/V$, the energy density $\langle H\rangle/V$ and the effective nucleon mass $m^{*}$ as functions of the chemical potential $\mu$ at fixed temperature $T$ for the RMF approach. The curves $1,2,3,4$ are obtained at $T=7,15$ MeV, $T=T_{c}$ and $T=22$ MeV, respectively. The solid lines $1,2$ are the results with the Maxwell construction. The symbol is the critical point and the dashed line is the phase diagram.} \label{a3}
\end{figure}

\begin{figure}[htbp]
\includegraphics[width=14.5cm]{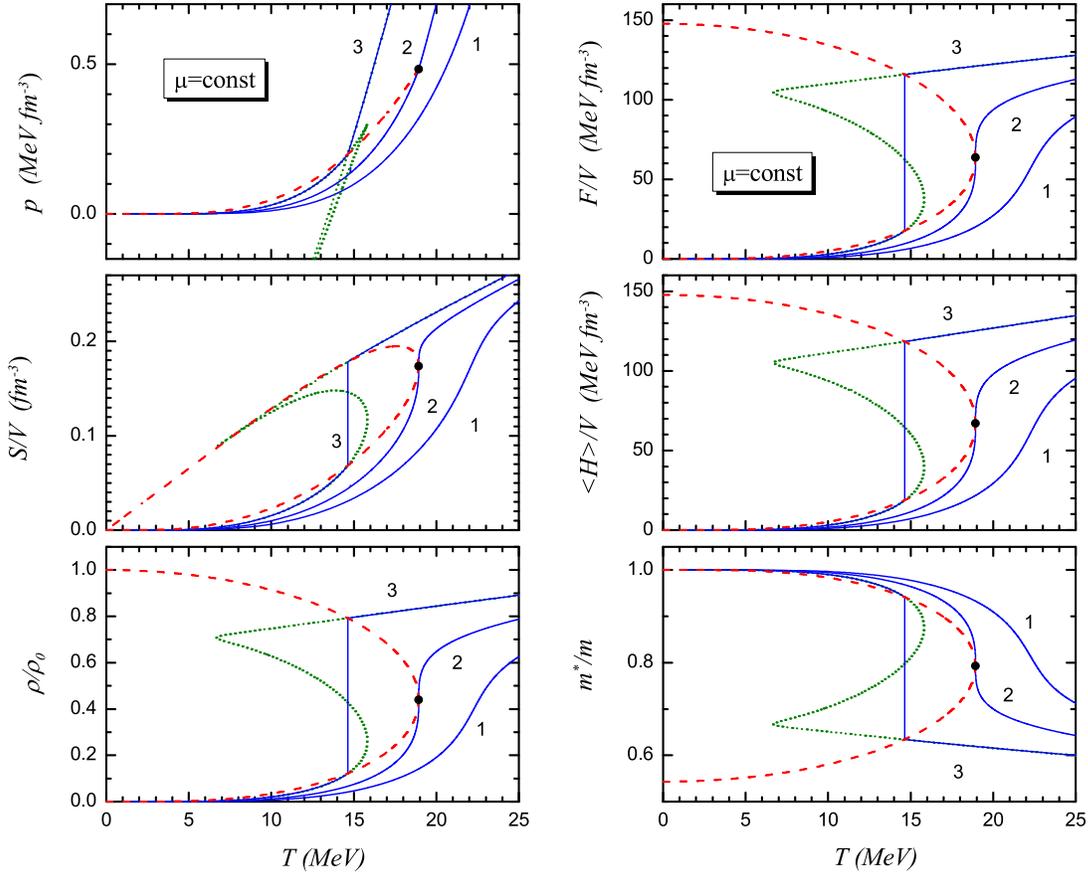}  \vspace{-0.3cm}
\caption{(Color online) Grand canonical ensemble $(T,\mu)$. The specific thermodynamical potential $p$ (the pressure), the entropy density $S/V$, the baryon density $\rho$, the density of free energy $F/V$, the energy density $\langle H\rangle/V$ and the effective nucleon mass $m^{*}$ as functions of the temperature $T$ at fixed chemical potential $\mu$ for the RMF approach. The curves $1,2,3$ are calculated for $\mu=905$ MeV, $\mu=\mu_{c}=910.033$ MeV and $\mu=914.95$ MeV, respectively. The continuous curve $3$ is the results with the Maxwell construction. Symbol depicts the critical point. The dashed line is the phase diagram.} \label{a4}
\end{figure}

The thermodynamic functions are multiple valued, as it can be seen in Figs. \ref{a1}, \ref{a3} and \ref{a4}, and therefore the stable states are obtained by the Maxwell construction, by excluding the loops--the areas A, B, and C in Fig. \ref{a1}. The self-intersection points $a$ on the curves $p(\mu)$ and $p(T)$ are found from the solutions of the equations
\begin{equation}\label{57a}
    p_{1}(\mu)=p_{3}(\mu), \qquad p_{1}(T)=p_{3}(T),
\end{equation}
where $p_{1}$ and $p_{3}$ are the values of the specific thermodynamical potential $p$ (the pressure) on the lines $1$ and $3$, respectively. These
solutions are the numbers $\mu^{*}(T)$ and $T^{*}(\mu)$. In this way, the loops A on the curves $p(\mu)$ and $p(T)$ in Figs. \ref{a1}(a) and \ref{a1}(b), respectively, are replaced by the points $a$, whereas the beckbending curves connecting the points $b$ and $c$ on the functions $\rho(\mu)$ and $\rho(T)$ (Figs. \ref{a1}c and \ref{a1}d) are replaced by straight lines. This definition of the Maxwell construction is equivalent with the equal-area definition for $p-v$ dependence \cite{Huang} as represented in the the bottom panels of Fig.~\ref{a1}.

The critical point, $(T_{c},\mu_{c},\rho_{c})$, is obtained from the equation
\begin{equation}\label{59a}
    \min\left(\frac{\partial\mu(\rho)}{\partial\rho}\right)=0
\end{equation}
and for the typical numerical parameters of the RMF model, with a nucleon mass, $m=939$ MeV, we find $T_{c}=18.927$ MeV, $\mu_{c}=910.033$ MeV and $\rho_{c}=0.441 \rho_{0}$.

Figure~\ref{a3} represents the behavior of $p$, $s$, $\rho$, $f$, $\varepsilon$ and $m^{*}$, as functions of the chemical potential $\mu$ at fixed temperature $T$; $f\equiv F/V$ is the density of free energy. For $T\geq T_{c}$, the thermodynamical variables $p(\mu)$, $\rho(\mu)$, $s(\mu)$, $\varepsilon(\mu)$, $f(\mu)$ and $m^{*}(\mu)$ are continuous, one-valued, monotonic and differentiable functions of $\mu$ (the line $3,4$ in all panels of Fig.~\ref{a3}). For $T<T_{c}$, the specific grand potential $p(\mu)$, with the Maxwell construction, is a piecewise smooth function. The point of phase transition, denoted by $\mu^*$, is a point of discontinuity of the first derivative of $p$ (the solid lines $1,2$ in Fig.~\ref{a3}). For $T<T_c$, with the Maxwell construction, $p$ is continuous and single-valued for all $\mu$, but its first order partial derivatives with respect to variables of state, namely $\rho(\mu)$ and  $s(\mu)$, are discontinuous at $\mu=\mu^{*}$. The jump of the entropy density at $\mu=\mu^*$ is related to the latent heat--$\lambda=T\delta s$. Moreover, $p$, $\rho$ and $s$, are strictly increasing functions of $\mu$.

Because of the small values of $\rho$ and the slow increase of $p$ with $\mu$, we say that the region $\mu\leq\mu^{*}$ corresponds to the gas phase of the nuclear matter.

In the region $\mu\geq\mu^{*}$ the baryon density is considerably larger and the presssure increases more rapidly with $\mu$, so we call this the liquid phase of the nuclear matter.

The baryon density, the entropy density, the density of free energy and the energy density have positive jump discontinuities throughout the phase separation at the point of phase transition, $\mu^{*}$, whereas the effective nucleon mass $m^{*}$ has a negative jump.

\begin{figure}
\includegraphics[width=13cm]{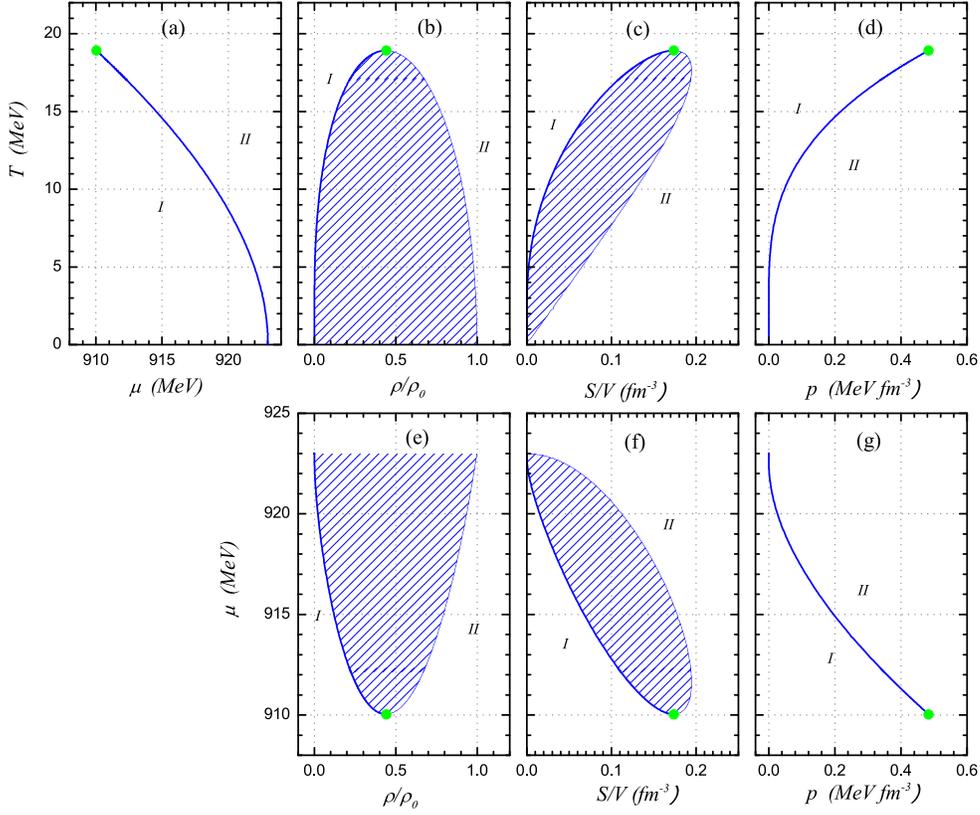} \vspace{-0.3cm}
\caption{(Color online) The phase diagrams $T-\mu$, $T-\rho$, $T-s$, $T-p$ (top panels) and $\mu-\rho$, $\mu-s$, $\mu-p$ (bottom panels) for the first order phase transition of the RMF approach. Roman numerals denote the homogeneous phases, the gas phase ($I$) and the liquid phase ($II$), the shaded areas correspond to the mixed phase, whereas the lines are the coexistence curves. Symbol depicts the critical point.} \label{a5}
\end{figure}

In Fig.~\ref{a4} we plot $p$, $s$, $\rho$, $f$, $\varepsilon$ and $m^{*}$ as functions of $T$, at fixed $\mu$. If $\mu\leq \mu_{c}$ or $\mu\geq \mu_{0}$ ($\mu_{0}$ is the chemical potential on the coexistence curve at $T=0$--see Fig.~\ref{a5}), the functions $p(T)$, $\rho(T)$, $s(T)$, $\varepsilon(T)$, $f(T)$ and $m^{*}(T)$ are continuous, single-valued, monotonic and differentiable for any $T$ (the lines $1$ and $2$ in Fig.~\ref{a4}). At fixed $\mu_{c}<\mu<\mu_{0}$, the function $p(T)$ with Maxwell construction is a piecewise smooth function with a point of discontinuity of the first order derivatives at $T=T^{*}$ (the line $3$ in Fig.~\ref{a4}). In this point $p(T)$ is continuous, but has a sharp corner (cusp). The first order partial derivatives of $p$ with respect to variables of state, namely $\rho(T)$ and $s(T)$, with Maxwell construction are single-valued piecewise continuous functions with jump discontinuities at $T=T^{*}$. Therefore, according to the Ehrenfest clasification for phase transitions, the system undergoes a first order phase transition at $T=T^*$, $\mu$ being fixed in the interval $\mu_{c}<\mu<\mu_{0}$.

The function $p(T)$, $\rho(T)$ and $s(T)$, are strictly increasing functions of $T$.  At $T\leq T^{*}$, the baryon density is small and the system is in the gas phase, whereas at $T\geq T^{*}$ the baryon density is large and the system is in the liquid phase. The entropy density, the baryon density, the density of free energy and the energy density have positive jump discontinuities throughout the phase separation at the point of phase transition $T^{*}$, whereas $m^{*}$ has a negative jump discontinuity.

The phase diagrams for the symmetric nuclear matter in the grand canonical ensemble $(T,\mu)$ in the RMF approach are depicted in Fig.~\ref{a5}. The phase diagram $T-\mu$ for the first order (liquid-gas) phase transition is represented by the continuous coexistence curve which crosses the $\mu$ axes at the point $(T=0,\mu=\mu_{0})$ and it is finished at the critical point $(T_{c},\mu_{c})$, where $\mu_{0}=m+\varepsilon_{b0}=923$ MeV. Along this curve, $\rho(T,\mu)$ and $s(T,\mu)$, are discontinuous. The phase diagrams $T-p$ and $\mu-p$ also are represented by the continuous coexistence curves because the function $p(T,\mu)$ is continuous in the points of phase transition. However, in the phase diagrams $T-\rho$, $T-s$, $\mu-\rho$, and $\mu-s$ we have coexistence regions instead of coexistence lines, because the variables $\rho$ and $s$ are undefined in the points of phase transition $(T^{*},\mu^{*})$. Note that the mixed phase is defined by the coexistence lines on the phase diagrams $T-\mu$, $T-p$ and the coexistence areas on the phase diagrams $T-\rho$, $T-s$.

In conclusion, the phase transition that appears in the nuclear matter in the RMF model in the grand canonical ensemble is accompanied by discontinuities in the first order derivatives of the thermodynamic potential $\Omega$ and therefore is a first order phase transition in the Ehrenfest classification~\cite{Ehrenfest}.

\subsection{Isobaric ensemble $(T,p,B)$}
Let us investigate the phase transition of the RMF model in the isobaric ensemble. The thermodynamical potential of the isobaric ensemble is the Gibbs  free energy $G=\Omega+p V+\mu B$ and its first order partial derivatives with respect to the variables of state, $(T,p,B)$, are the entropy, the volume and the chemical potential,
\begin{equation}\label{81a}
    S=-\left(\frac{\partial G}{\partial T} \right)_{pB}, \quad  V=\left(\frac{\partial G}{\partial p} \right)_{TB},
    \quad \mu= \left(\frac{\partial G}{\partial B} \right)_{Tp}.
\end{equation}
The potential $G$ is proportional to $B$, and therefore the specific thermodynamic potential is
\begin{equation}\label{53ab}
\tilde{g}(T,p) \equiv G(T,p,B)/B = \mu(T,p).
\end{equation}
The first order partial derivatives of the specific thermodynamical potential $\mu(T,p)$ with respect to the variables of state $(T,p)$ can be written as
\begin{equation}\label{53ac}
   \tilde{s}(T,p)=-\left(\frac{\partial \mu}{\partial T} \right)_{p}, \qquad  v(T,p)=\left(\frac{\partial \mu}{\partial p} \right)_{T},
\end{equation}
where $\tilde{s}=S/B$ is the entropy per nucleon and $v=V/B$ is the specific volume. They satisfy the relations
\begin{equation}\label{53ad}
  d\mu = -\tilde{s}dT+v dp,  \qquad  T\tilde{s}=\tilde{\varepsilon}+pv-\mu,
\end{equation}
where $\tilde{\varepsilon}=\langle H\rangle/B$ is the energy per nucleon.

\begin{figure}[t]
\includegraphics[width=14.5cm]{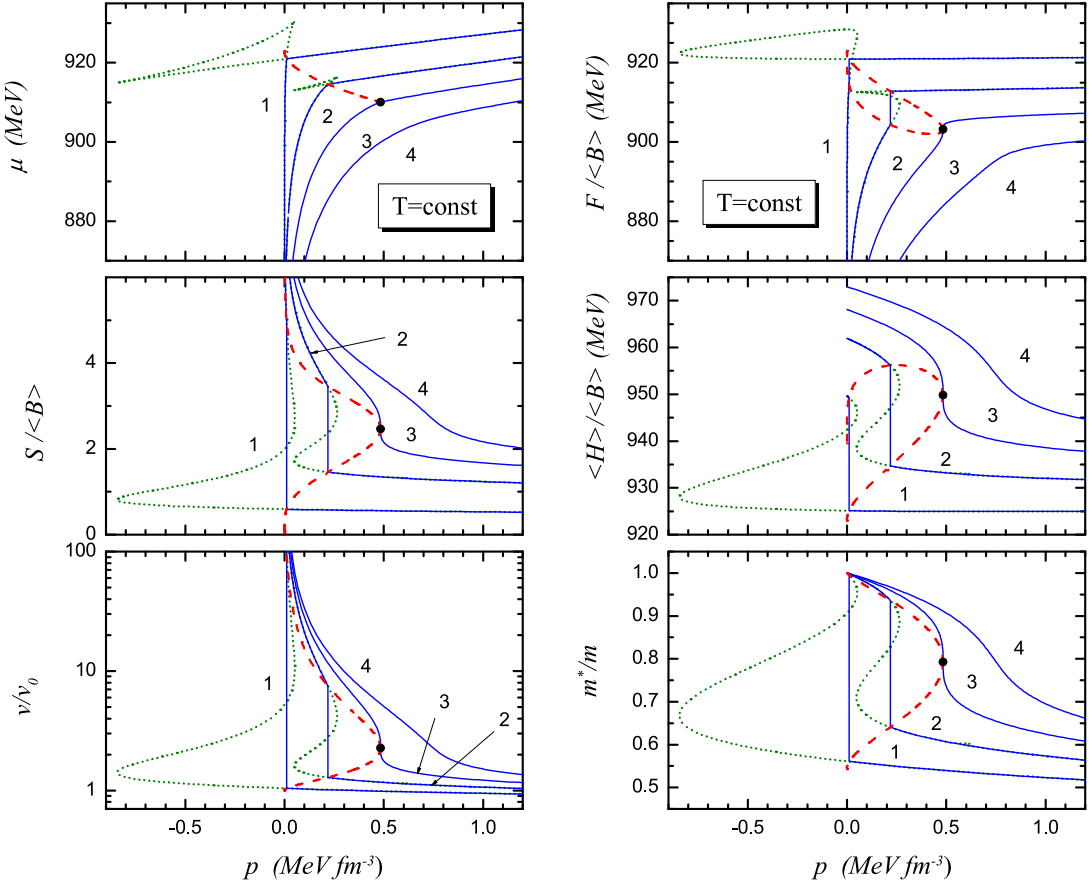} \vspace{-0.3cm}
\caption{(Color online) Isobaric ensemble $(T,p)$. The Gibbs free energy per nucleon $\mu$ (the chemical potential), the entropy per nucleon $\tilde{s}$, the specific volume $v$, the free energy per nucleon $\tilde{f}$, the mean energy per nucleon, $\tilde{\varepsilon}$ and the effective nucleon mass $m^{*}$ as functions of the pressure $p$ at fixed temperature $T$ for the RMF model. The curves 1, 2, 3, and 4 are obtained for $T=7,15$ MeV, $T=T_{c}$ and $T=22$ MeV, respectively. The solid lines 1  and 2 correspond to the Maxwell construction. The symbol is the critical point and the dashed line is the phase diagram.} \label{a6}
\end{figure}

In Fig. \ref{a6} we plot the functions $\mu$, $\tilde{s}$, $v$, $\tilde{f}$, $\tilde{\varepsilon}$ and $m^*/m$ as functions of $p$, at fixed $T$ for the RMF approach; $\tilde{f}=F/B$ is the free energy per nucleon. For $T\geq T_{c}$, the functions $\mu(p)$, $v(p)$, $\tilde{s}(p)$, $\tilde{\varepsilon}(p)$, $\tilde{f}(p)$ and $m^{*}(p)$ are continuous, one-valued, monotonic and differentiable for any $p$ (the lines $3,4$ in Fig.~\ref{a6}). For $T<T_{c}$, $\mu(p)$ is still a continuous function, but its first order derivatives are discontinuous at the point of phase transition, $p^{*}$ (the solid lines 1 and 2 in Fig.~\ref{a6}). The first order partial derivatives of the thermodynamic potential with respect to variables of state, i.e. the functions $v(p)$ and $\tilde{s}(p)$, with the Maxwell construction, have jump discontinuities at $p=p^*$. Therefore in the Ehrenfest classification this is also a first order, liquid-gas, phase transition.

The jump of the entropy per nucleon at the phase transition pressure, $p^{*}$, is related to the latent heat. The functions $\mu(p)$ and $\tilde{f}(p)$ are increasing with $p$, whereas $v(p)$, $\tilde{s}(p)$, $\tilde{\varepsilon}(p)$ and $m^{*}(p)$ are decreasing. In the gas phase, $p<p^{*}$, the specific volume, $v(p)$, the entropy per nucleon, $\tilde{s}(p)$, and the energy per nucleon, $\tilde{\varepsilon}(p)$, have large values and the reduced effective nucleon mass is close to one ($m^{*}(p)/m\sim 1$). In the liquid phase ($p>p^{*}$), the functions $v(p)$, $\tilde{s}(p)$, $\tilde{\varepsilon}(p)$ and $m^{*}(p)/m$ take smaller values. The functions $v(p)$, $\tilde{s}(p)$, $\tilde{\varepsilon}(p)$ and $m^{*}(p)$ have negative jumps at the phase transition pressure, $p^{*}$, whereas the free energy per nucleon $\tilde{f}(p)$ has the positive jump discontinuity.

\begin{figure}[htb]
\includegraphics[width=14.5cm]{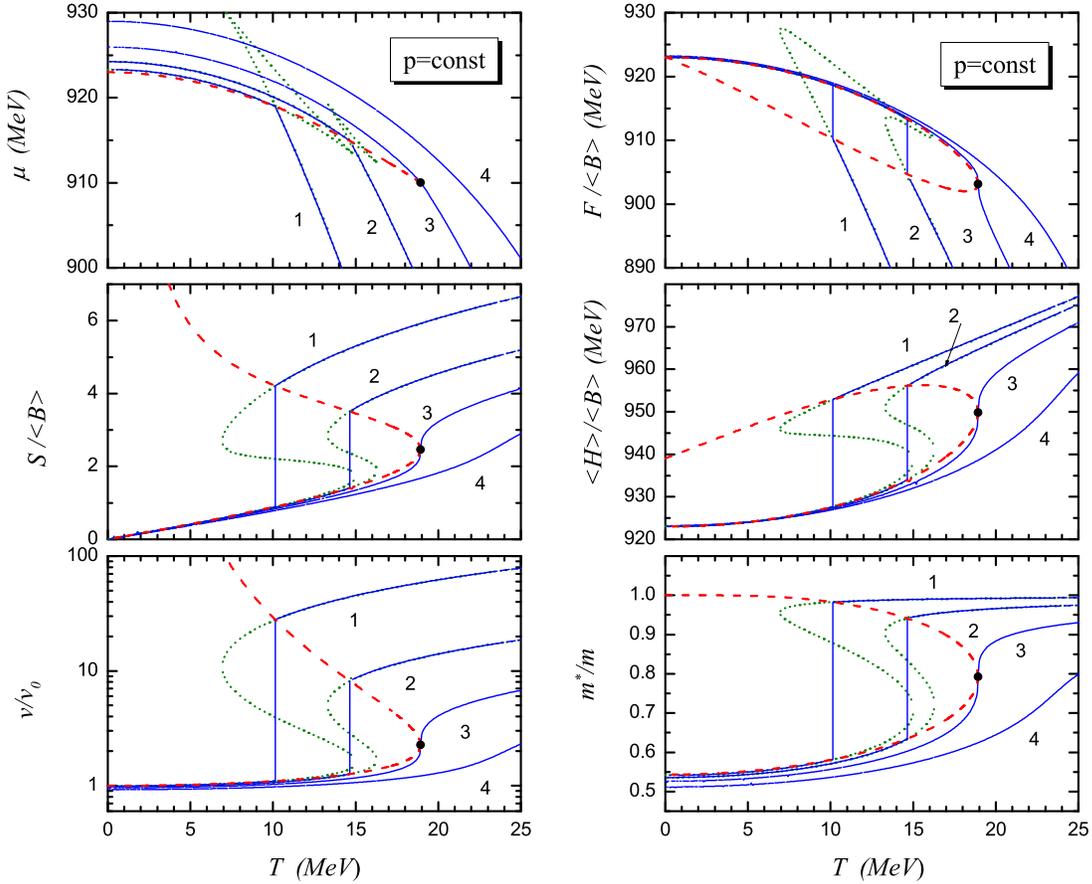} \vspace{-0.3cm}
\caption{(Color online) Isobaric ensemble $(T,p)$. The Gibbs free energy per nucleon, $\mu$ (the chemical potential), the entropy per nucleon, $\tilde{s}$, the specific volume, $v$, the free energy per nucleon, $\tilde{f}$, the mean energy per nucleon, $\tilde{\varepsilon}$, and the effective nucleon mass, $m^{*}$, as functions of $T$ at fixed $p$ for the RMF approach. The curves 1, 2, 3, and 4 are obtained at the pressure $p=0.05$ MeVfm$^{-3}$,$0.2$ MeV fm$^{-3}$, $p=p_{c}=0.484$ MeV fm$^{-3}$ and $p=1.0$ MeV fm$^{-3}$, respectively. The solid lines 1 and 2 are the results of the Maxwell construction. The symbol is the critical point at $p_{c}=0.484$ MeV fm$^{-3}$. The dashed line is the phase diagram.} \label{a7}
\end{figure}

Figure~\ref{a7} presents the behavior of $\mu$, $\tilde{s}$, $v$, $\tilde{f}$, $\tilde{\varepsilon}$ and $m^*/m$, as functions of temperature $T$ at fixed $p$ for the RMF approach. At the pressures higher than $p_c$, the functions $\mu(T)$, $v(T)$, $\tilde{s}(T)$, $\tilde{\varepsilon}(T)$, $\tilde{f}(T)$ and $m^{*}(T)$ are continuous, single-valued, monotonic and differentiable functions on all of $T$ (the lines 3 and 4 in Fig.~\ref{a7}).

At pressures below $p_{c}$, the specific Gibbs potential, $\mu(T)$, with the Maxwell construction, is a piecewise smooth function with a singularity at the phase transition temperature, $T^{*}$ (the solid lines 1 and 2 in Fig.~\ref{a7}). The first order partial derivatives of $\mu$ with respect to variables of state, i.e. the functions $v(T)$ and $\tilde{s}(T)$, are single-valued piecewise continuous and smooth function for all $T>0$, except in $T=T^*$, where they have jump discontinuities. Also the functions $\tilde{f}$, $\tilde{\varepsilon}$ and $m^{*}$ have jump discontinuities at the point of the phase transition.

The discontinuities of the quantities $\tilde{s}(T)$, $v(T)$, $\tilde{\varepsilon}(T)$ and $m^{*}(T)$ are positive, whereas the discontinuity of $\tilde{f}(T)$ is negative. Therefore in this ensemble also the phase transition is of the first order, according to the Ehrenfest clasification.

The functions $\mu(T)$ and $\tilde{f}(T)$ are monotonically decreasing, whereas the functions $v(T)$, $\tilde{s}(T)$, $\tilde{\varepsilon}(T)$ and $m^{*}(T)$ are monotonically increasing. In the gas phase, which correspond to a temperature $T>T^{*}$, the specific volume $v(T)$, the entropy per nucleon $\tilde{s}(T)$ and the energy per nucleon $\tilde{\varepsilon}(T)$ have large values, whereas the reduced effective nucleon mass is close to one ($m^{*}(T)/m\sim 1$). In the liquid phase, $T<T^{*}$, the functions $v(T)$, $\tilde{s}(T)$, $\tilde{\varepsilon}(T)$ and $m^{*}(T)$ take smaller values.

Summarizing, we showed that in the isobaric ensemble the thermodynamic potential, $\mu(T,p)$ (the chemical potential), is a continuous function, with piecewise continuous derivatives. The first order derivatives with respect to $T$ and $p$, which are the entropy per nucleon $\tilde{s}$ and the specific volume $v$, respectively, have jump discontinuities at the phase transition. The phase diagram in the coordinates $T-p$ is represented by a continuous coexistence curve (Fig. \ref{a5}d), whereas in the coordinates $T-v$ and $T-\tilde{s}$ is represented by the coexistence regions because the variables $v$ and $\tilde{s}$ are undefined in the points of phase transition at fixed values of $T$ and $p$.

Being accompanied by a jump discontinuity of the first order derivatives of the thermodynamic potential $G$, the phase transition of the RMF model in the isobaric ensemble is a first order phase transition in the Ehrenfest classification \cite{Ehrenfest} and, therefore, it is a liquid-gas type~\cite{Papon,Yeomans}.


\subsection{Canonical ensemble $(T,V,B)$}
Let us investigate the phase transition of the RMF model in the canonical ensemble. The thermodynamic potential of the canonical ensemble is the Helmholtz free energy, $F(T,V,B)=\Omega+\mu B$.  The first order partial derivatives of $F$ with respect to the variables of state, $(T,V,B)$, are 
\begin{equation}\label{82a}
    S=-\left(\frac{\partial F}{\partial T} \right)_{VB}, \quad  p=-\left(\frac{\partial F}{\partial V} \right)_{TB},
    \quad \mu= \left(\frac{\partial F}{\partial B} \right)_{TV}.
\end{equation}
Since $F$ is a homogeneous function in the variables $V$ and $B$, in the following we shall work with the specific thermodynamical potential (the free energy per nucleon)
\begin{equation}\label{83ab}
\tilde{f}(T,v) = F(T,V,B)/B.
\end{equation}
The first order partial derivatives of the specific thermodynamical potential $\tilde{f}(T,v)$ with respect to the variables of state $(T,v)$ can be written as
\begin{equation}\label{83ac}
   \tilde{s}(T,v)=-\left(\frac{\partial \tilde{f}}{\partial T} \right)_{v}, \qquad  p(T,v)=-\left(\frac{\partial \tilde{f}}{\partial v} \right)_{T}.
\end{equation}
Due to the homogeneity property of the thermodynamical potential (\ref{83ab}) all specific functions of variables of state of the canonical ensemble are intensive quantities.  These quantities satisfy the differential equation for $\tilde{f}$ and the Euler theorem
\begin{equation}\label{83ad}
  d\tilde{f} = -\tilde{s}dT-pdv,  \qquad  T\tilde{s}=\tilde{\varepsilon}+pv-\mu.
\end{equation}
To obtain the functions of the canonical ensemble we change the variables
of state of the grand canonical ensemble $(T,\mu)$ to the variables $(T,v)$ together with the Legendre transformation for the thermodynamical potential, $\tilde{f}=pv-\mu$.

\begin{figure}
\includegraphics[width=14.5cm]{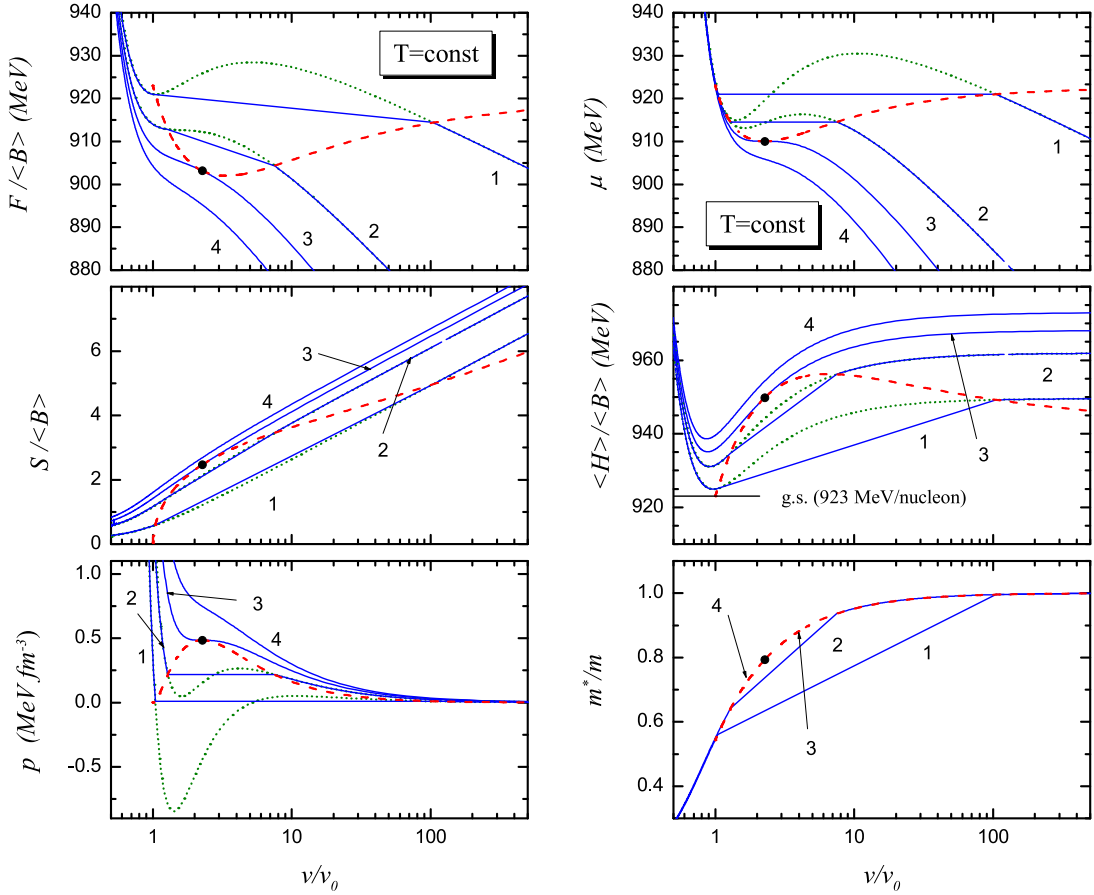}  \vspace{-0.4cm}
\caption{(Color online) Canonical ensemble $(T,v)$. The free energy per nucleon $\tilde{f}$, the entropy per nucleon $\tilde{s}$, the pressure $p$, the chemical potential $\mu$, the mean energy per nucleon $\tilde{\varepsilon}$ and the effective nucleon mass $m^{*}$ as functions of the specific volume $v$ at fixed temperature $T$ for the RMF approach. The curves $1,2,3,4$ are obtained at $T=7,15$ MeV, $T=T_{c}$ and $T=22$ MeV, respectively. The continuous curves $1,2$ are the results with the Maxwell construction. The symbol depicts the critical point and the dashed line is the phase diagram.} \label{a8}
\end{figure}
Figure~\ref{a8} presents the behavior of the specific thermodynamical potential $\tilde{f}$, the entropy per nucleon $\tilde{s}$, the pressure $p$, the chemical potential $\mu$, the energy per nucleon $\tilde{\varepsilon}$ and the effective nucleon mass $m^{*}$ as functions of the specific volume $v$ at fixed temperature $T$ for the RMF approach in the canonical ensemble. At temperature $T\geq T_{c}$, the exact thermodynamical quantities
$\tilde{f}(v)$, $\tilde{s}(v)$, $p(v)$, $\mu(v)$, $\tilde{\varepsilon}(v)$ and $m^{*}(v)$ are continuous, one-valued, monotonic and differentiable functions on all of $v$ (the lines $3,4$ in Fig.~\ref{a8}). At the temperature $T<T_{c}$ the specific Helmholtz potential $\tilde{f}(v)$ with the Maxwell construction, its first derivatives, the entropy per nucleon $\tilde{s}(v)$ and the pressure $p(v)$, and the quantities $\mu(v)$, $\tilde{\varepsilon}(v)$ and $m^{*}(v)$ are characterized by the linear changes in a closed  interval $v_{II}<v<v_{I}$, where $v_{I}$ $(v_{II})$ are the specific volume $v$ in the gas (liquid) phase across the phase boundary at fixed temperature $T$ (see the solid lines $1,2$ in Fig.~\ref{a8}). The chemical potential $\mu(v)$ and the pressure $p(v)$ (isotherms) have the plateau as functions of $v$ at fixed temperature $T$ because they take the constant values in the points of phase transition. The function $\tilde{f}(v)$ has the linear decrease, but the functions $\tilde{s}(v)$, $\tilde{\varepsilon}(v)$ and $m^{*}(v)$ have linear growth in the region of mixed phase, $v_{II}<v<v_{I}$. In the liquid phase, at $v<v_{II}$ and $T<T_{c}$, the entropy per nucleon $\tilde{s}(v)$ and the effective nucleon mass $m^{*}$ decrease with decreasing of $v$, but the functions $p(v)$, $\mu(v)$ and $\tilde{f}(v)$ increase. The energy per nucleon $\tilde{\varepsilon}$ achieves his minimum at $v=v_{0}$. In the gas phase, at $v>v_{I}$ and $T<T_{c}$, the entropy per nucleon $\tilde{s}(v)$, the energy per nucleon $\tilde{\varepsilon}$ and the effective nucleon mass $m^{*}$ increase with increasing $v$, but the functions $p(v)$, $\mu(v)$ and $\tilde{f}(v)$ decrease.

\begin{figure}
\includegraphics[width=14.5cm]{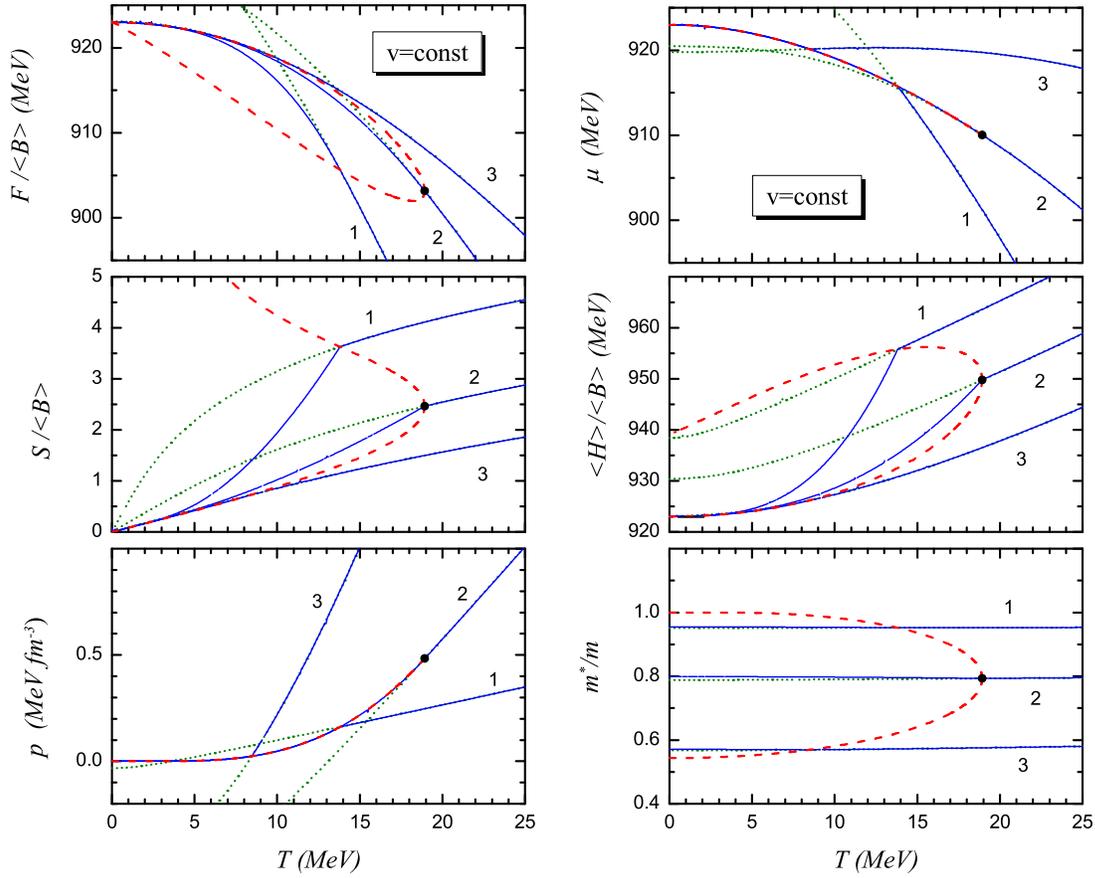}\vspace{-0.3cm}
\caption{(Color online) Canonical ensemble $(T,v)$.  The free energy per nucleon $\tilde{f}$, the entropy per nucleon $\tilde{s}$, the pressure $p$, the chemical potential $\mu$, the mean energy per nucleon $\tilde{\varepsilon}$ and the effective nucleon mass $m^{*}$ as functions of the temperature $T$ for the fixed  specific volume $v$ (the baryon density $\rho=1/v$) for the RMF approach. The curves $1,2,3$ depict the results at the specific volume $v/v_{0}=10$ ($\rho/\rho_{0}=0.1$), $v=v_{c}$ ($\rho=\rho_{c}$) and $v/v_{0}=1.064$ ($\rho/\rho_{0}=0.94$), respectively. The solid lines $1,2,3$ are the results with the Maxwell construction. Symbol is the critical point at $v_{c}=2.268 v_{0}$. The dashed line is the phase diagram.} \label{a9}
\end{figure}
Figure~\ref{a9} presents the behavior of the specific thermodynamical potential $\tilde{f}$, the entropy per nucleon $\tilde{s}$, the pressure $p$, the chemical potential $\mu$, the energy per nucleon $\tilde{\varepsilon}$ and the effective nucleon mass $m^{*}$ as functions of the temperature $T$ at fixed specific volume $v$ for the RMF approach in the canonical ensemble. The Maxwell construction at fixed values of $v$ for any function $A(T,v)$ is defined by the phenomenological equation
\begin{equation}\label{55a}
  A = A_{I}+\alpha (A_{II}-A_{I}), \quad  \alpha = \frac{\rho-\rho_{I}}{\rho_{II}-\rho_{I}},
\end{equation}
where $A_{I}$ $(A_{II})$ and $\rho_{I}$ $(\rho_{II})$ are the thermodynamical quantity $A(T,\mu)$ and the baryon density $\rho(T,\mu)$ in the gas (liquid) phase across the phase boundary in the grand canonical ensemble. Then, the function $A(T,v)$ of the canonical ensemble is obtained from the Eq.~(\ref{55a}) by changing the variables of state $(T,\mu)$ into the variables of state $(T,v)$. At the fixed specific volume from the interval $v_{0}\leq v <\infty$, the specific Helmholtz potential, $\tilde{f}(T)$, with the Maxwell construction, its first derivatives, the entropy per nucleon $\tilde{s}(T)$ and the pressure $p(T)$, and the quantities $\mu(T)$, $\tilde{\varepsilon}(T)$ and $m^{*}(T)$ begin to deviate from their exact values at the temperatures $T<T_{s}$, where $T_{s}$ is temperature of the crossing point on the coexistence curve. The first derivatives, $\tilde{s}(T)$ and $p(T)$, and the quantities $\mu(T)$ and $\tilde{\varepsilon}(T)$ are single-valued continuous broken-line functions and they have sharp corners (cusps) at the temperature $T=T_{s}$. This point determines the transition of the system from the mixed phase at $T<T_{s}$ to the gas (liquid) phase at $T>T_{s}$. Therefore, at fixed specific volume $v$ the system can be only in two phase: mixed and gas (liquid). The functions $\mu(T)$ and $\tilde{f}(T)$ are strictly decreasing functions on a set $I$, but the functions $\tilde{s}(T)$, $p(T)$ and $\tilde{\varepsilon}(T)$ are strictly increasing functions on a set $I$. The functions $\mu(T)$ and $p(T)$ in the mixed phase at $T<T_{s}$ coincide with their coexistence curves because they take the constant values in the points of phase transition.

Summarizing, we have obtained that in the canonical ensemble the first order phase transition of the nuclear liquid-gas type associated with the Gibbs free energy is defined by the linear changes of the specific thermodynamical potential $\tilde{f}$ and its first order partial derivatives, the entropy per nucleon $\tilde{s}$ and the pressure $p$, as functions of the specific volume $v$ at fixed temperature $T$. Otherwise, it is also defined by the curvilinearly varying of the specific thermodynamical potential, the entropy per nucleon and the pressure, as functions of the temperature $T$ at fixed specific volume $v$ and by the sharp corners of the first derivatives $\tilde{s}$ and $p$ in the points where they cross the coexistence curves. The phase diagram $T-v$ is depicted by the coexistence aria instead of the coexistence line.

\subsection{Caloric curve and the equation of state}

Let us investigate the caloric curve, i.e. the dependence of the temperature $T$ on the energy $E$ of the system, for the phase transition of the RMF model in the grand canonical, isobaric, canonical and microcanonical ensembles. The thermodynamical potential of the microcanonical ensemble is the entropy $S(E,V,B)$. The first order partial derivatives of the thermodynamical potential $S$ with respect to the variables of state $(E,V,B)$ are the temperature, the pressure and the chemical potential
\begin{equation}\label{84a}
    \frac{1}{T}=\left(\frac{\partial S}{\partial E} \right)_{VB}, \quad  \frac{p}{T}=\left(\frac{\partial S}{\partial V} \right)_{EB},
    \quad \frac{\mu}{T}= -\left(\frac{\partial S}{\partial B} \right)_{EV}.
\end{equation}
The Legendre transformation for the thermodynamical potential is the equation $TS=E-\Omega-\mu B$.

The thermodynamical potential of the microcanonical ensemble, the entropy $S$, is a homogeneous function of first degree of the extensive variables of state $V,B$
\begin{equation}\label{85ab}
S(E,V,B) = B \tilde{s}(\tilde{\varepsilon},v),
\end{equation}
where $\tilde{s}(\tilde{\varepsilon},v)$ is the specific thermodynamical potential, $v=V/B$ and $\tilde{\varepsilon}=E/B$. Therefore, the variable $B$ is excluded and the microcanonical ensemble is described by the specific thermodynamical potential  $\tilde{s}(\tilde{\varepsilon},v)$ as a function of the intensive variables of state $(\tilde{\varepsilon},v)$. Then, the first order partial derivatives of the specific thermodynamical potential $\tilde{s}(\tilde{\varepsilon},v)$ with respect to the variables of state $(\tilde{\varepsilon},v)$ can be written as
\begin{equation}\label{85ac}
   T(\tilde{\varepsilon},v)=\left(\frac{\partial \tilde{s}}{\partial \tilde{\varepsilon}} \right)^{-1}_{v}, \qquad
    p(\tilde{\varepsilon},v)=T\left(\frac{\partial \tilde{s}}{\partial v} \right)_{\tilde{\varepsilon}},
\end{equation}
Due to the homogeneity property of the thermodynamical potential (\ref{85ab}) all specific functions of variables of state of the microcanonical ensemble are intensive quantities. These quantities satisfy the differential equation for $\tilde{s}$ and the Euler theorem
\begin{equation}\label{85ad}
 T d\tilde{s} = d\tilde{\varepsilon} + pdv,  \qquad  T\tilde{s}=\tilde{\varepsilon}+pv-\mu,
\end{equation}
To obtain the functions of the microcanonical ensemble we change the variables of state of the grand canonical ensemble $(T,\mu)$ to the variables $(\tilde{\varepsilon},v)$ together with the Legendre transformation for the thermodynamical potential, $\tilde{s}=(\tilde{\varepsilon}+pv-\mu)/T$.

\begin{figure}
\includegraphics[width=15cm]{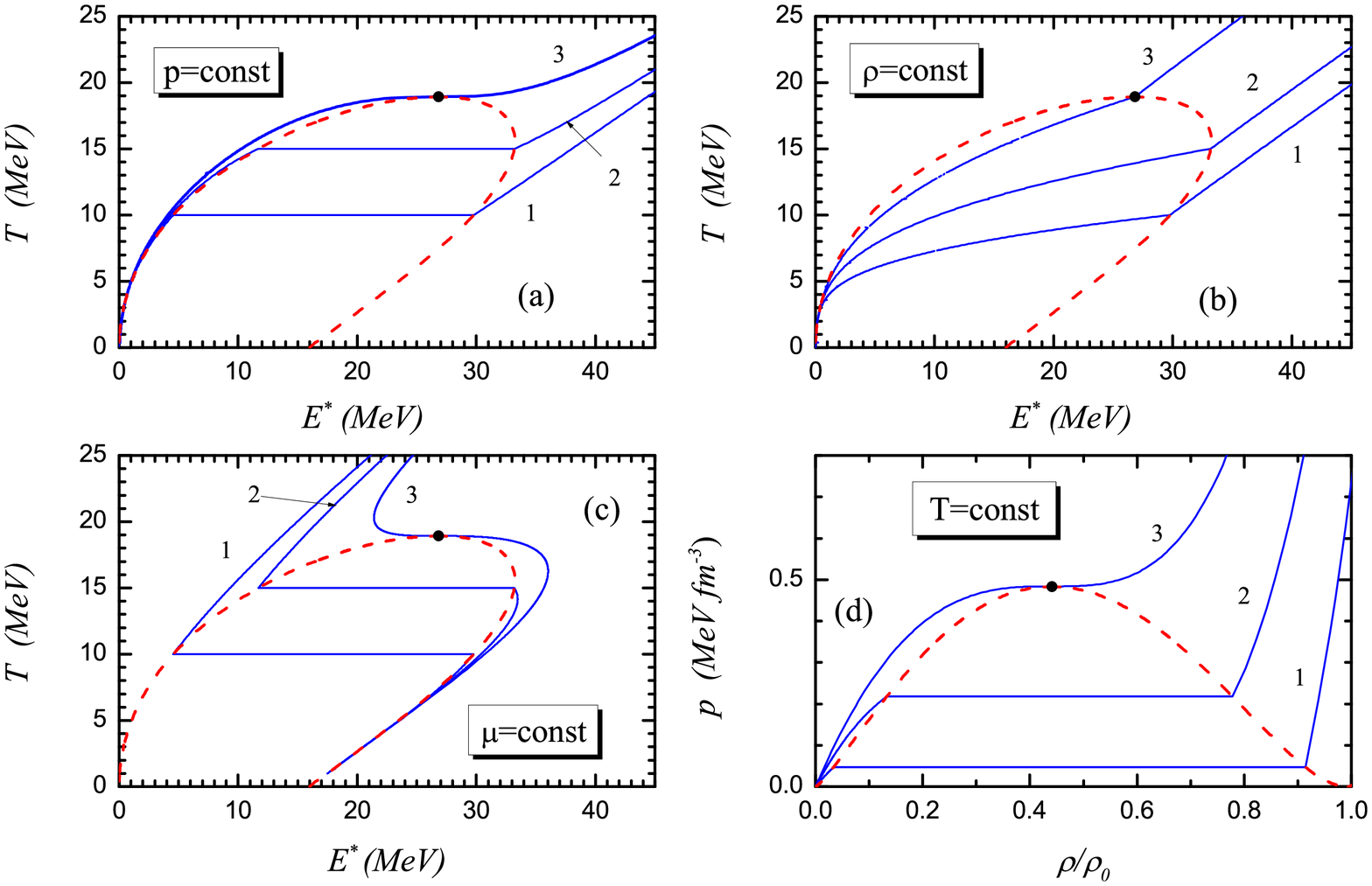} \vspace{-0.3cm}
\caption{(Color online) The temperature $T$ as function of the excitation energy per nucleon $E^{*}$ (the caloric curve) for the RMF model in the different statistical ensembles at three points of phase transition (lines $1,2,3$): (a) the isobaric ensemble at fixed pressure $p$, (b) the canonical and microcanonical ensembles at fixed baryon density $\rho$ and (c) the grand canonical ensemble at fixed chemical potential $\mu$. (d) The isotherms in the canonical and isobaric ensembles at fixed temperature $T$ (solid lines) in the same points of phase transition. Symbol is the critical point and the dotted line is the phase diagram. For details see the text and the Table~\ref{tab1}.}   \label{a10}
\end{figure}
\begin{table}[htbp]
\caption{The parameters of the three points of phase transition for the RMF model} \label{tab1}
\begin{tabular}{ c  p{1cm} c p{2cm} c p{2cm} c p{2cm} c p{2cm} c}
  \hline  \hline
    &  & $T$     & & $\mu$     & &  $\rho/\rho_{0}$    & &  $p$           & &  $v/v_{0}$   \\
    &  & MeV     & & MeV       & &                     & &  MeV fm$^{-3}$ & &              \\ \hline
 1  &  & 10      & & 919.066   & &  0.034              & &  0.047         & &  29.256       \\
 2  &  & 15      & & 914.557   & &  0.133              & &  0.217         & &  7.513        \\
 3  &  & $T_{c}$ & & $\mu_{c}$ & & $\rho_{c}/\rho_{0}$ & &  $p_{c}$       & &  $v_{c}/v_{0}$ \\ \hline \hline
\end{tabular}
\end{table}

Figure~\ref{a10} presents the caloric curve in the isobaric, canonical, microcanonical and grand canonical ensembles, and the equation of state or the isotherms, i.e. the dependence $p-\rho$ at fixed temperature $T$, in the canonical and isobaric ensembles for the nuclear liquid-gas phase transition of the RMF model. The excitation energy per nucleon is defined by the equation, $E^{*}=\tilde{\varepsilon}(T)-\tilde{\varepsilon}(0)$, where $\tilde{\varepsilon}(0)=923$ MeV is the energy per nucleon in the ground state. The lines $1,2,3$ in Fig.~\ref{a10} were calculated for three points of phase transition. In Table~\ref{tab1}, the parameters of these points are summarized. In the point of phase transition at temperature $T=T^{*}$ the excitation energy per nucleon $E^{*}$ in the caloric curve has a jump discontinuity in the isobaric and grand canonical ensembles at fixed pressure $p$ and at fixed chemical potential $\mu$, respectively. Otherwise, in the canonical and microcanonical ensembles at fixed specific volume $v$ or baryon density $\rho$ the excitation energy per nucleon $E^{*}$ in the caloric curve does not have a jump discontinuity. The curve $T(E^{*})$ is a single-valued continuous broken-line function and has a sharp corner at the temperature $T=T_{s}$, where it crosses the coexistence curve. In the mixed phase at $T<T_{s}$ the caloric curve $T(E^{*})$ in the canonical and microcanonical ensembles is a continuous increasing function.

As $T$ increases, the system in the isobaric ensemble at fixed pressure $p$ exhibits the phase transition from the liquid phase to the gas phase through an intermediate mixed phase created in the point of phase transition at, temperature $T=T^{*}$. In the grand canonical ensemble at fixed chemical potential $\mu$ the system exhibits the phase transition from the gas phase to the liquid phase also through the mixed phase. But, in the canonical and microcanonical ensembles at fixed baryon density $\rho$, as $T$ increases, only the transition from the mixed phase to the gas (liquid) phase takes place. Moreover, at small temperatures and $0<\rho/\rho_{0}<1$ the nuclear matter in the canonical and microcanonical ensembles is situated in the mixed phase. In particular, the liquid-gas phase transition is characterized by the jump discontinuity of the baryon density $\rho$ or specific volume $v$ in the points of phase transition, i.e. it is defined by the plateau in the isotherms in the isobaric and canonical ensembles at fixed temperature $T$. See Fig.~\ref{a10}. The uncertainty of $\rho$ in the points of liquid-gas phase transition excludes the discontinuity of the energy in the canonical and microcanonical ensembles at fixed values of $\rho$, because for the one fixed value of the baryon density at constant temperature we enable to choose only one value of the energy. Therefore, the caloric curve for the liquid-gas phase transition in these two ensembles do not contain the plateau.

Summarizing, we have found that for the nuclear liquid-gas phase transition of the RMF model the energy in the caloric curve is discontinuous in the isobaric and the grand canonical ensembles at fixed values of the pressure and the chemical potential, respectively, and it is continuous, i.e. it has no plateau, in the canonical and the microcanonical ensembles at fixed values of the baryon density. However, the baryon density in the equation of state (the isotherm) is discontinuous in the isobaric and the canonical ensembles at fixed values of the temperature. For one of the variants of the SMM the similar results were obtained in~\cite{Aguiar06}. Thus, the general criterion for the nuclear liquid-gas phase transition, i.e. the first order phase transition associated with the Gibbs free energy $G$, in the canonical ensemble requires that the baryon density in the isotherms should be discontinuous at fixed values of the temperature and the energy in the caloric curves should be continuous, i.e. it should not have plateau, at fixed values of the baryon density (the baryon charge and the volume). Note that if for the certain physical system the energy in the caloric curve in the canonical and the microcanonical ensembles at constant values of the variables of state of the canonical ensemble is discontinuous, then its phase transition is a first order phase transition associated with the free energy $F$, because the discontinuity of the energy $E$, when $F$ is continuous, is related to the discontinuity of the entropy, $E=F+TS$, which is the first derivative of the potential $F$~\cite{Stanley}. We should also mention that all results of this work for the RMF model concerning the properties of the first order phase transition of the liquid-gas type associated with the Gibbs free energy $G$ are in agreement with the principles of the general theory of phase transitions~\cite{Papon,Yeomans,Chomaz02}.

\section{Conclusions}\label{Concl}
The first order phase transition for the RMF model was investigated on the basis of the method of the thermodynamical potentials and their first derivatives in different statistical ensembles. The main thermodynamical properties of this phase transition were found by using the Maxwell construction in the framework of the grand canonical, canonical and isobaric ensembles. It was established that the first order phase transition of the RMF model is the phase transition of the nuclear liquid-gas type which can be associated with the Gibbs free energy $G$ and its properties totally satisfy the requirements of the general theory of phase transitions. Indeed, we have found that in the isobaric ensemble the Gibbs free energy per nucleon (the chemical potential) for the RMF model is the piecewise smooth function and its first order partial derivatives with respect to variables of state $(T,p)$, i.e., the entropy per nucleon and the specific volume, are the piecewise continuous functions. In the points of phase transition the chemical potential is a continuous function which has a cusp both as a function of $p$ at fixed $T$ and as a function of $T$ at fixed $p$ and the first order partial derivatives of the chemical potential with respect to variables of state, the entropy per nucleon and the specific volume, have jump discontinuities.

Also, we have revealed that for the RMF model in the grand canonical ensemble the specific grand potential (the pressure) is the piecewise smooth function and its first order partial derivatives with respect to variables of state $(T,\mu)$, i.e. the entropy density and the baryon density, are the piecewise continuous functions. In the points of phase transition the pressure is a continuous function which has a sharp corner (cusp) both as a function of $\mu$ at fixed $T$ and as a function of $T$ at fixed $\mu$, and the first derivatives, the entropy density and the baryon density, have jump discontinuities. However, in the canonical ensemble the definition of the nuclear liquid-gas phase transition appears unable to meet the criteria established for the grand canonical and isobaric ensembles. The first order derivatives of the Helmholtz free energy per nucleon with respect to variables of state, the entropy per nucleon and the pressure, as functions of specific volume (baryon density) at fixed temperature vary linearly in the region of phase transition, but as functions of the temperature at fixed specific volume vary smoothly and have sharp corners in the points where they cross the coexistence curves. The first derivatives of the free energy have no jump discontinuities in the canonical ensemble. Thus, the nuclear liquid-gas phase transition in the RMF model is characterized by the jump discontinuities of the baryon density (the specific volume) and the entropy density (the entropy per nucleon) at constant values of temperature, chemical potential and pressure. This implies that for the nuclear liquid-gas phase transition of this model the phase diagrams $T-\mu$ and $T-p$ are represented by coexistence lines, however, the phase diagrams $T-s$ ($T-\tilde{s}$) and $T-\rho$ ($T-v$) are depicted by the coexistence areas.

The caloric curve and the equation of state for the RMF model were calculated. It was established that the energy in the caloric curve for the nuclear liquid-gas phase transition of the RMF model is discontinuous in the isobaric and the grand canonical ensembles at fixed values of the pressure and the chemical potential, respectively, and it is continuous in the canonical and microcanonical ensembles at fixed values of baryon density or the specific volume. However, the baryon density in the isotherms is discontinuous in the isobaric and the canonical ensembles at fixed values of the temperature. The general criterion for the nuclear liquid-gas phase transition in the canonical ensemble was identified. It states that the baryon density in the isotherms should be discontinuous at fixed values of the temperature and the energy in the caloric curves should be continuous at fixed values of the baryon density. It should be also mentioned that the obtained results for the first order phase transition of the RMF model are in total agreement with the principles of the general theory of phase transitions.

{\bf Acknowledgments:} This work was supported in part by the joint research project of JINR and IFIN-HH, protocol N~4063 and the RFBR grant~08-02-01003-a. I acknowledge valuable remarks and fruitful discussions with A.S.~Sorin and D.V.~Anghel. I also thank S.~Cojocaru and K.A.~Bugaev for their comments.



\begin{thebibliography} {xxxx}

\bibitem{Yagi} K.~Yagi, T.~Hatsuda, Y.~Miake, Quark-gluon plasma. From big bang to little bang (Cambridge University Press, 2005).

\bibitem{Bondorf} J.P.~Bondorf, A.S.~Botvina, A.S.~Iljinov, I.N.~Mishustin, K.~Sneppen, Phys. Rep. 257 (1995) 133.

\bibitem{Steiner05} A.W.~Steiner, M.~Prakash, J.M.~Lattimer, P.J.~Ellis, Phys.Rep. 411 (2005) 325.



\bibitem{Mishustin06} I.N.~Mishustin, Eur. Phys. J. A 30 (2006) 311.

\bibitem{Bondorf85} J.P.~Bondorf, R.~Donangelo, I.N.~Mishustin, H.~Schulz, Nucl. Phys. A 444 (1985) 460.

\bibitem{Pochodzalla} J.~Pochodzalla et al., Phys. Rev. Lett. 75 (1995) 1040.

\bibitem{Hauger} J.A.~Hauger et al., Phys. Rev. Lett. 77 (1996) 235.

\bibitem{Ma97} Y.G.~Ma et al., Phys. Lett. B 390 (1997) 41.



\bibitem{Bondorf98} J.P.~Bondorf, A.S.~Botvina, I.N.~Mishustin, Phys. Rev. C 58 (1998) R27.

\bibitem{Gross90} D.H.E.~Gross, Rep. Prog. Phys. 53 (1990) 605.

\bibitem{Das2003} C.B.~Das, S.~Das~Gupta, A.Z.~Mekjian, Phys. Rev. C 68 (2003) 031601 (R).

\bibitem{Raduta02} A.H.~Raduta, A.R.~Raduta, Nucl. Phys. A 703 (2002) 876.

\bibitem{Scharenberg01} R.P.~Scharenberg et al., Phys. Rev. C 64 (2001) 054602.

\bibitem{Parvan00}  A.S.~Parvan, V.D.~Toneev, M.~P{\l}oszajczak, Nucl. Phys. A 676 (2000) 409.



\bibitem{Elliott00} J.B.~Elliott, A.S.~Hirsch, Phys. Rev. C 61 (2000) 054605.

\bibitem{Samaddar04} S.K.~Samaddar, J.N.~De, S.~Shlomo, Phys. Rev. C 69 (2004) 064615.

\bibitem{Aguiar06} C.E.~Aguiar, R.~Donangelo, S.R.~Souza, Phys. Rev. C 73 (2006) 024613.

\bibitem{De07} J.N.~De, S.K.~Samaddar, Phys. Rev. C 76 (2007) 044607.

\bibitem{DasGupta01} S.~Das Gupta, A.Z.~Mekjian, M.B.~Tsang, Adv. Nucl. Phys. 26 (2001) 89.

\bibitem{Parvan99} A.S.~Parvan, V.D.~Toneev, K.K.~Gudima, Yad. Fiz. 62 (1999) 1593 [Phys. At. Nucl. 62 (1999) 1497].



\bibitem{Campa09} A.~Campa, T.~Dauxois, S.~Ruffo, Phys. Rep. 480 (2009) 57.

\bibitem{Huller94} A.~H\"uller, Z. Phys. B 93 (1994) 401.

\bibitem{Ota01} S.~Ota, S.B.~Ota, Phys. Lett. A 285 (2001) 247; Int. J. Mod. Phys. B 21 (2007) 3591.

\bibitem{Gross97} D.H.E.~Gross, Phys. Rep. 279 (1997) 119.

\bibitem{Lee96} K.-C.~Lee, Phys. Rev. E 53 (1996) 6558.

\bibitem{Chomaz99} P.~Chomaz, F.~Gulminelli, Nucl. Phys. A 647 (1999) 153.

\bibitem{Chomaz02} P.~Chomaz, F.~Gulminelli, Lect. Notes Phys. 602 (2002) 68.

\bibitem{Barre02} J.~Barr\'e, D.~Mukamel, S. Ruffo, Lect. Notes Phys. 602 (2002) 45.

\bibitem{Mulken01} O.~M\"ulken, H.~Stamerjohanus, P.~Borrmann, Phys. Rev. E 64 (2001) 047105.


\bibitem{Ehrenfest} P.~Ehrenfest, Commun. Kamerlingh Omnes Lab. Univ. Leiden Suppl. 75b (1933).

\bibitem{Papon} P.~Papon, J.~Leblond, P.H.E.~Meijer, The physics of phase transitions (Springer-Verlag, Berlin, 2006).

\bibitem{Yeomans} J.M.~Yeomans, Statistical mechanics of phase transitions (Clarendon Press, Oxford, 1992).

\bibitem{Stanley} H.E.~Stanley, Introduction to phase transitions and critical phenomena (Clarendon Press, Oxford, 1971).



\bibitem{Landau} L.D.~Landau, E.M.~Lifshitz, Statistical physics (Pergamon, Oxford, 1989).

\bibitem{Yang} C.N.~Yang, T.D.~Lee, Phys. Rev. 87 (1952) 404; \\
   T.D.~Lee, C.N.~Yang, Phys. Rev. 87 (1952) 410.

\bibitem{Fisher} M.E.~Fisher, in Lectures in Theoretical Physics, Vol. 7c (University of Colorado Press, Boulder, 1965).

\bibitem{Lee} K.-C.~Lee, Phys. Rev. Lett. 73 (1994) 2801.

\bibitem{Chomaz1} P.~Chomaz, F.~Gulminelli, V.~Duflot, Phys. Rev. E 64 (2001) 046114.

\bibitem{Chomaz2} P.~Chomaz, F.~Gulminelli, Physica A 305 (2002) 330; Physica A 330 (2003) 451.

\bibitem{Gross} D.H.E.~Gross, Physica E 29 (2005) 251; Physica A 365 (2006) 138.

\bibitem{Gorenstein} M.I.~Gorenstein, M.~Ga\'zdzicki, W.~Greiner, Phys. Rev. C 72 (2005) 024909.

\bibitem{Bugaev} K.A.~Bugaev, M.I.~Gorenstein, I.N.~Mishustin, W.~Greiner, Phys. Rev. C 62 (2000) 044320.



\bibitem{Walecka74} J.D.~Walecka, Ann. Phys. 83 (1974) 491; \\
F.E.~Serr, J.D.~Walecka, Phys. Lett. B 79 (1978) 10.

\bibitem{Serot86} B.D.~Serot, J.D.~Walecka, Adv.Nucl.Phys. 16 (1986) 1; Int. J. Mod. Phys. E 6 (1997) 515.

\bibitem{Silva} J.B.~Silva, O.~Louren\c{c}o, A.~Delfino, J.S.~S\'{a} Martins, M.~Dutra, Phys. Lett. B 664 (2008) 246.

\bibitem{Avancini04} S.S.~Avancini, L.~Brito, D.P.~Menezes, C.~Provid\^{e}ncia, Phys. Rev. C 70 (2004) 015203.

\bibitem{Avancini06} S.S.~Avancini, L.~Brito, Ph.~Chomaz, D.P.~Menezes, C.~Provid\^{e}ncia, Phys. Rev. C 74 (2006) 024317.

\bibitem{Ducoin08} C.~Ducoin, C.~Provid\^{e}ncia, A.M.~Santos, L.~Brito, Ph.~Chomaz, Phys. Rev. C 78 (2008) 055801.

\bibitem{Ayik11} S.~Ayik, O.~Yilmaz, F.~Acar, B.~Danisman, N.~Er, A.~Gokalp, Nucl. Phys. A 859 (2011) 73.

\bibitem{Ducoin06} C.~Ducoin, Ph.~Chomaz, F.~Gulminelli, Nucl. Phys. A 771 (2006) 68.



\bibitem{Goodman84} A.L.~Goodman, J.I.~Kapusta, A.Z.~Mekjian, Phys. Rev. C 30 (1984) 851.

\bibitem{Huang} K.~Huang, Statistical mechanics (Wiley, New York, 1987).



\bibitem{Okun} L.B.~Okun, Leptons and quarks (North-Holland, Amsterdam, 1982).

\bibitem{Anghel} D.V.~Anghel, A.S.~Parvan, A.S.~Khvorostukhin, Physica A 391 (2012) 2313.



\end{thebibliography}
\end{document}